\documentclass{article}

\PassOptionsToPackage{numbers, compress}{natbib}

\usepackage[preprint]{neurips_2024}

\usepackage[utf8]{inputenc} 
\usepackage[T1]{fontenc}    
\usepackage{hyperref}       
\usepackage{url}            
\usepackage{booktabs}       
\usepackage{amsfonts}       
\usepackage{nicefrac}       
\usepackage{microtype}      
\usepackage{xcolor}         
\usepackage{graphicx}       
\usepackage{listings}         
\usepackage{caption}

\title{HEnRY: A Multi-Agent System Framework for Multi-Domain Contexts}

\author{%
  Emmanuele Lacavalla\thanks{ORCID: \url{https://orcid.org/0000-0002-2204-1212}} \\
  \And
  Shuyi Yang \\
  \And
  Riccardo Crupi \\
  \And
  Joseph E. Gonzalez \\
}

\begin{document}

\maketitle

\begin{abstract}
  This project, named \textbf{HEnRY}, aims to introduce a Multi-Agent System (MAS) into Intesa Sanpaolo. The name HEnRY summarizes the project's core principles: the \textbf{H}ierarchical organization of agents in a layered structure for efficient resource management; \textbf{E}fficie\textbf{n}t optimization of resources and operations to enhance overall performance; \textbf{R}eactive ability of agents to quickly respond to environmental stimuli; and \textbf{Y}ielding adaptability and flexibility of agents to handle unexpected situations. The discussion covers two distinct research paths: the first focuses on the system architecture, and the second on the collaboration between agents. This work is not limited to the specific structure of the Intesa Sanpaolo context; instead, it leverages existing research in MAS to introduce a new solution. Since Intesa Sanpaolo is organized according to a model that aligns with international corporate governance best practices, this approach could also be relevant to similar scenarios.
\end{abstract}

\section{Introduction}
Assist employees in performing their daily activities is not only a concern of knowledge. In an complex context, an assistant, in addition to providing answers to questions about company knowledge, it can also perform actions within the information system by carrying out certain tasks. The peculiarity of this work lies in the environment, as the solution is applied across different independent domains. The context of this work pertains to Intesa Sanpaolo, a prominent European banking group and the leading financial institution in Italy across all business sectors. The Intesa Sanpaolo Group\footnote{https://group.intesasanpaolo.com/en/about-us} serves 13.6 million customers domestically and maintains a strategic international presence, providing services to an additional 7.4 million customers worldwide.

\begin{figure*}[ht]
  \centering
  \includegraphics[width=\textwidth]{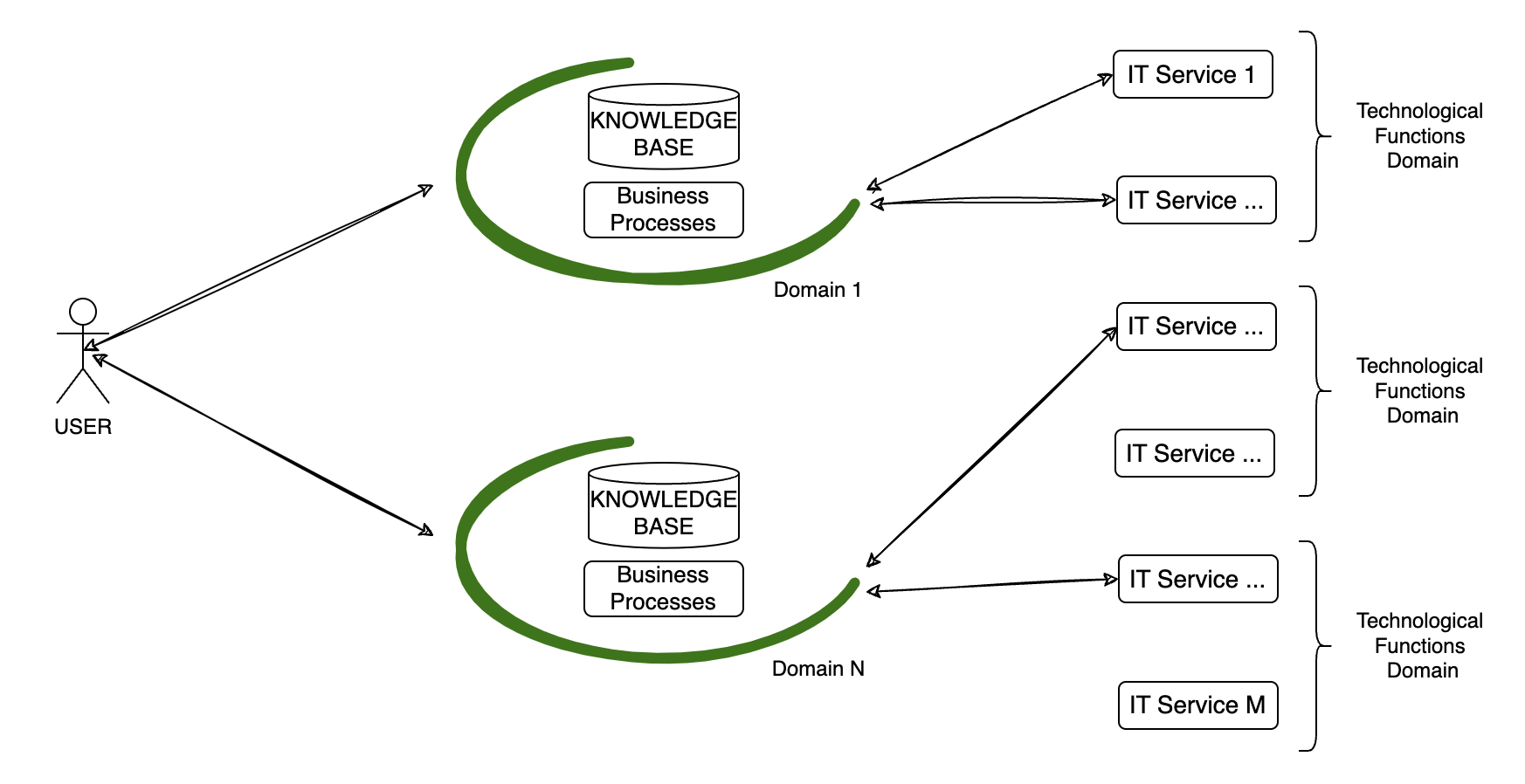}
  \caption{The image depicts how domains work within Intesa Sanpaolo. The user interacts with domains, each contain business processes and a knowledge base. These domains rely on IT services provided by specific technological domains, which are managed by domain experts to ensure proper operation and support of internal processes.}
  \label{fig:isp_structure}
\end{figure*}

In a complex environment like Intesa Sanpaolo, knowledge management can be challenging. As showed in the Figure \ref{fig:isp_structure}, the company structure has a tree-like, hierarchical organization. This condition leads to the creation of domains within the knowledge base. Additionally, the strict regulations in this banking context significantly divide information, creating very specific visibility cones for each type of domain. One of the challenges was to segment domains knowledge according to the end user. This situation arises from the principle of least privilege, which is applied to every single user. Indeed, each person can access information relevant to their role. 
Another aspect relates to the architecture, particularly the topology of the system. As matter of, within each domain, there are not only knowledge and procedures but also technological functions that provide services necessary for the domain's operation. This leads to a segmentation of domain-specific technological experts who can operate within their context. This means that each domain function is managed by a domain expert with proper tools and processes. Given the complexity of this scenario, this work was preceded by the creation of a chatbot for a single domain. This solution will be described in appendix \ref{appendix_a1}. Based on this initial experience, the HEnRY project at Intesa Sanpaolo is being structured. The aims of this work is to introduce a MAS across multiple domains.
Introducing a MAS requires careful consideration and study of certain components to make it easily integrable and, therefore, pervasive. This aspect also requires various cybersecurity measures. The nature of these distributed applications is essential to comply with all necessary regulations while ensuring security and high reliability at the same time.

The organizational structure of Intesa Sanpaolo and the actual solution will be described in Section \ref{contex}. Section \ref{review} will recap the current literature. Section \ref{kc} will present the key challenges. The proposed solution will be detailed in Section \ref{solution}. Finally, Section \ref{conclusions} will discuss the work and present the conclusions of the study.
\section{Context and Issue Outline}
Intesa Sanpaolo is organized according to a model that aligns with international corporate governance best practices.\footnote{https://group.intesasanpaolo.com/en/governance/central-structures} As a result, this work does not focus merely on the structure of the Intesa Sanpaolo context; instead, it leverages existing research in MAS to propose an innovative solution in this field. The organizational structure \cite{ISP} is customer-oriented, with seven main Divisions targeting specific market segments. Alongside these Divisions, the central Governance Areas are responsible for guidance, coordination, and control. These Governance Areas cover a range of responsibilities, including human resources, administration and control, finance, legal matters, social responsibility and sustainability, technology and innovation, communication and external relations, cost, risk, and credit management. Each of these areas has specific knowledge related to its concepts, procedures, and responsibilities. This collective body of knowledge constitutes the definition of \textbf{domains}. Typically, each domain of knowledge has its own channel as a touchpoint (i.e., the point of access for the knowledge in that area), and this could represent an issue for those who need to interact across different domains when necessary.

The version of the chatbot (based on a RAG approach) developed prior to this work is tailored to a single domain. However, this approach led to the creation of silos for each new domain, resulting in several issues that represent the starting points for this work:

\begin{itemize}
    \item \textbf{Scalability and Governance of the Solution}: The key problem here is how the system performs as the number of domains increases. If not addressed, the amount of data, the level of services, and the functionality could vary significantly and may not integrate seamlessly. Additionally, without proper governance, managing and arbitrating between different domains could become problematic, as each domain has specific owners, making the sharing of a single chatbot challenging.
    \item \textbf{Efficiency and Cloud Providers}: There is a risk of developing similar software for different domains in different ways, which could lead to inefficiencies in leveraging common features and services. Moreover, managing the interchangeability of cloud services across different providers could become sluggish without a unique framework, leading to extensive work required to recustomize all components of the chatbots.
    \item \textbf{User Access}: The central problem is how to create an omnichannel application that allows interaction with all agents uniformly. If not solved, users may be forced to use different portals to interact with assistants, which could pose significant risks in data expositions.
\end{itemize}
\label{contex}
\section{Literature Review}
\citet{stone2000multiagent} emphasize the importance of AS in scenarios where multiple individuals or departments have distinct goals and proprietary information. In these cases, MAS plays a crucial role in managing interactions between these entities. When departments need to model their internal operations, it is often challenging for them to agree on a single system that meets everyone’s needs. Each department requires a system tailored to its unique capabilities and priorities. A solution could be to allow each department to develop agents that represent its specific goals and interests. These agents can then be integrated into a MAS using specialized techniques. From this perspective, \citet{dorri2018multi} describe MAS as a platform where autonomous entities, known as agents, work together to solve tasks. Typically, these agents can learn, make independent decisions, and interact with their peers or the environment. MAS is often modeled as a graph, with middle agents used to reduce the overhead involved in finding the appropriate agent.

\citet{zhang2023building} show that recent research demonstrates the use of Large Language Models (LLMs) to drive agents in single-agent tasks through zero-shot prompting for instruction-following or few-shot prompting for more complex, long-horizon tasks. However, the creation of cooperative agents, that can work with other agents or humans in decentralized settings, remains a challenging and an underexplored area.

\citet{liu2023dynamic} argue that the field of LLM-agent collaborations requires a systematic framework to improve generalizability, efficiency, and performance. Ideally, MAS should dynamically adapt their composition to the specific domain of a query with minimal supervision. Although there have been some attempts, a comprehensive framework has yet to be developed. Their work addresses this gap by introducing a framework for LLM-agent collaboration on complex tasks. \citet{packer2023memgpt} point out the limitations of LLMs due to their restricted context windows, which hinder their effectiveness in tasks such as extended conversations and document analysis. They suggest virtual context management, as a solution to these limitations. This technique involves a system that intelligently manages different storage tiers to effectively extend the usable context within the LLM's limited context window.
\label{review}

\section{Key Challenges}
\label{kc}

This work, building on the previous experience described in Appendix \ref{appendix_a1}, aims to outline an evolutionary path from a single-domain chatbot to a MAS based on multiple domains, addressing the following questions: how to set up, what architecture to use, how to enable communication between agents in a multi-domain context structured like Intesa Sanpaolo, and how to make the approach non-invasive and reliable for the group?

The proposed solution aims to provide a scalable approach based on heterogeneous agents that can be easily deployed in a multi-domain context. The tools allow the creation of a network of agents, which can be managed collaboratively across various domains. When a task is assigned to the MAS, it is decomposed into all the necessary steps, which are solved by agents distributed across different domains. This focuses on an important point: each part of the problem is managed by its own domain of expertise, in accordance with established rules. Additionally, the agents within each domain can be heterogeneous, enabling the use of the most suitable LLMs for the domain's specific purpose.

Another crucial element is the intrinsic traceability within the system. For each request, it is possible to identify which agents were involved, ensuring accountability for the activation of the MAS network nodes. Furthermore, the proposed solution is based on a new context, whereas the projects discussed in Section \ref{review} is based on different or deeply specialized aspects, such as code generation.

\section{Proposed Solution and Forward-Looking Vision}
\label{solution}
This session will describe the proposal for the MAS of the HEnRY project and its research areas. In the repository \href{https://github.com/2mmanu/henry}{github.com/2mmanu/henry}, you can find a Kubernetes application capable of instantiating a MAS. An example is provided in Appendix \ref{appendix_b1}. The agents used in the MAS are based on agentBUDDY \cite{agentBUDDY}, a Python package that encapsulates the tools needed for the MAS and the system's agent abstraction. These two components are developed as part of this work.

\begin{figure*}[ht]
  \centering
  \includegraphics[width=\textwidth]{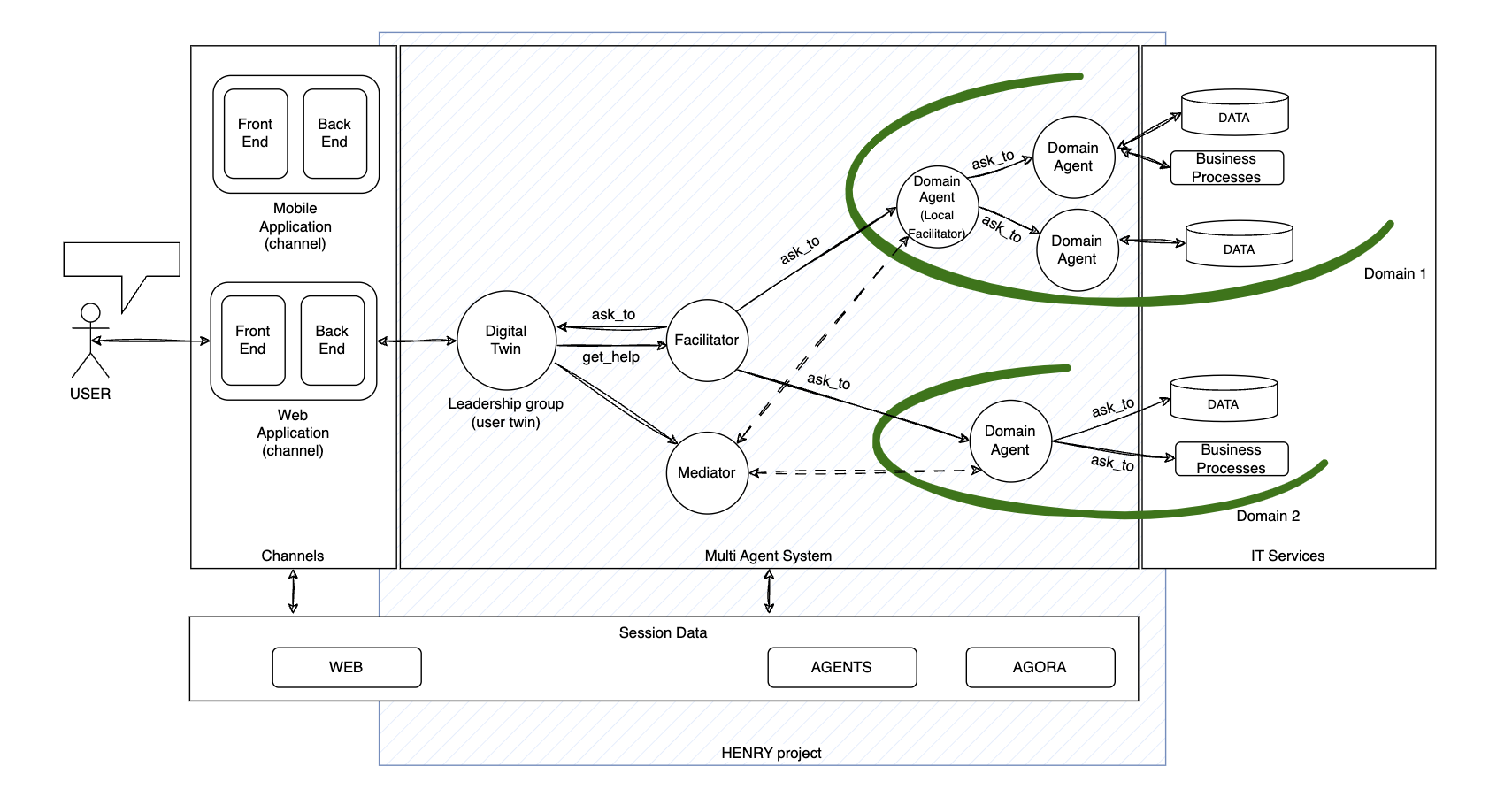}
  \caption{The system includes an agent called digital twin, designed to be a comprehensive assistant for the user offering complete customization. The facilitator, being aware of all domains, collaborates with the digital twin to solve problems across various areas. The digital twin can integrate useful information for the facilitator when further details are required. The domain agents are distributed and aligning the company hierarchical structure with that of the multi-agent system, ensuring both maintainability and security. As concern security, the compartmentalization of roles is effective, allowing each agent to verify the conditions necessary to carry out an operation. All agents share a session data service, where web and agents track operations, while agora facilitates cross-domain collaboration through an ephemeral agent called the mediator, who ensures control and security during access to inter-domain services.}
  \label{fig:proposed_solution}
\end{figure*}

As showed in Figure \ref{fig:proposed_solution}, the solution features a specific architecture with four types of agents: the digital-twin, the facilitator, the mediator, and the domain agent. In the MAS, there can be one or more instances of each type of agent.

\subsection{Digital Twin Concept}

The digital-twin represents the agent that instantiates the human user in the MAS. This agent has persistent memory and can extend its functionality by requesting support from a facilitator or mediator agent. Thus, the digital-twin can act as a leader within the MAS and request services from the system. As the digital entity of the user, this agent receives information related to its user or retrieves it from its memory based on historical data. Moreover, this agent can humanize the interaction with the MAS. For example, in case of local operational issues within the domains, the digital-twin can explain to the user what is happening without using technical or incomprehensible information, and the agent itself can store the request and resubmit it later, notifying the end user of the final result.

\subsection{Domain Agent Concept}

The domain agent is a specific agent for domains that can perform a particular role. These functionalities are possible because the agent has access to a knowledge base or services within the information system. In this way, the agent performs very simple and specific tasks, making it modular and controllable regarding its actions towards the information system. In a domain, there can be one or more domain agents. Each domain agent has a parent. A parent assumes the role of a facilitator, as it groups different domain agents. Thus, it will be possible to access various basic services through a single access point. This is configured as a tree from an architecture point of view. When a domain agent joins the network, it notifies its parent nodes with a syntax of its capabilities to help them send appropriate requests. This process continues across all layers, allowing each node to grasp the key topic of its respective section without needing all the details.

\subsection{Facilitator Concept}

The primary function of this agent is to answer questions based on knowledge distributed across the domains through its child agents. A digital twin can receive assistance from a facilitator. This process divides questions into specific queries for the appropriate child domain agent and so forth. Additionally, if further information is required, the facilitator can request integrations from the digital twin without returning to the user with information already known to the system. This approach helps avoid requiring information that the user might consider obvious. An example of the agent is provided in Appendix \ref{appendix_a3}.

\subsection{Mediator Concept}
\label{fsm-mediator}

The mediator can set up an asynchronous job to solve a complex problem with the goal of performing an action or creating a digital resource by instantiating a group across the available domains.
The mediator works as an ephemeral agent with four stages. In the first stage, it receives a request and prepares an environment for addressing it, including creating a shared context called agora, recruiting or creating relevant agents, and providing feedback to the digital twin. In the second stage, the mediator collects initial solutions from all agents within the shared context. The third stage involves initiating a parallel discussion where agents refine their solutions. Finally, in the fourth stage, the mediator gathers the final solutions from the agents and publishes them in the shared area with the digital twin, facilitating discussion with the end user. An example of the agent can be found in Appendix \ref{appendix_a3}.
\section{Conclusions}
The initial phase of Project HEnRY involves establishing the project's discovery stage. The project's current maturity is not yet consolidated as it remains in the simulation phase. Additionally, the mentioned tools are still under development. The next steps of the project will be to create a extended baseline to evaluate the effectiveness of the upcoming design phases in an industrial environment. Project with such complexity, raise various questions, such as: "When should a human intervene? At which level?", "How to optimize the prompts used for agent collaboration?", or "How to verify whether the proposed solution is a pattern or an anti-pattern?"

This work marks the initial phase of the HEnRY project, aiming to establish a framework for the MAS solution. The project's structure includes two distinct research paths:

\textbf{Macro Viewpoint}: This path is engage in to defining the structure, topology, communication protocols, and components of the global MAS.

\textbf{Micro Viewpoint}: This path focuses on the collaboration of agents to address business procedures and common issues within the company.

The macro viewpoint ensures common integration of the system, while the micro viewpoint is crucial for effective task decomposition and collaboration, enabling the completion of designed tasks. The dual-path approach could facilitate both system-wide coherence and detailed agent interactions, paving the way for further development and refinement in the project's subsequent stages.
\label{conclusions}



\medskip

{
\small
\bibliographystyle{apalike}
\bibliography{main}

\begin{thebibliography}{}

\bibitem[2mmanu, 2024]{agentBUDDY}
2mmanu (2024).
\newblock agentbuddy: A companion for your multi-agent systems.
\newblock \url{https://github.com/2mmanu/agentBUDDY}.
\newblock GitHub repository. Accessed 28 July 2024.

\bibitem[Dorri et~al., 2018]{dorri2018multi}
Dorri, A., Kanhere, S.~S., and Jurdak, R. (2018).
\newblock Multi-agent systems: A survey.
\newblock {\em Ieee Access}, 6:28573--28593.

\bibitem[Es et~al., 2023]{es2023ragas}
Es, S., James, J., Espinosa-Anke, L., and Schockaert, S. (2023).
\newblock Ragas: Automated evaluation of retrieval augmented generation.
\newblock {\em arXiv preprint arXiv:2309.15217}.

\bibitem[{Intesa Sanpaolo}, 2024]{ISP}
{Intesa Sanpaolo} (2024).
\newblock Organizational structure.
\newblock \url{https://group.intesasanpaolo.com/en/about-us/organisational-structure}.
\newblock Accessed: 2024-07-9.

\bibitem[Liu et~al., 2023]{liu2023dynamic}
Liu, Z., Zhang, Y., Li, P., Liu, Y., and Yang, D. (2023).
\newblock Dynamic llm-agent network: An llm-agent collaboration framework with agent team optimization.
\newblock {\em arXiv preprint arXiv:2310.02170}.

\bibitem[Packer et~al., 2023]{packer2023memgpt}
Packer, C., Fang, V., Patil, S.~G., Lin, K., Wooders, S., and Gonzalez, J.~E. (2023).
\newblock Memgpt: Towards llms as operating systems.
\newblock {\em arXiv preprint arXiv:2310.08560}.

\bibitem[Stone and Veloso, 2000]{stone2000multiagent}
Stone, P. and Veloso, M. (2000).
\newblock Multiagent systems: A survey from a machine learning perspective.
\newblock {\em Autonomous Robots}, 8:345--383.

\bibitem[Zhang et~al., 2023]{zhang2023building}
Zhang, H., Du, W., Shan, J., Zhou, Q., Du, Y., Tenenbaum, J.~B., Shu, T., and Gan, C. (2023).
\newblock Building cooperative embodied agents modularly with large language models.
\newblock {\em arXiv preprint arXiv:2307.02485}.

\end{thebibliography}
}

\section*{Social Impact Statement}
This project aims to introduce a MAS by addressing the challenge of instantiating multiple domains in a financial/industrial environment. At present, the work does not have significant social impacts, as it is still in the early stages. However, as the project progresses and is consolidated in an industrial setting, potential broader impacts will need to be considered. This includes addressing ethical aspects, privacy concerns, security measures, and mechanisms for monitoring system learning and fairness. Exploring these areas will be crucial for understanding the future societal consequences of the system.

\appendix

\section{Appendix: Detailed Insights}
\subsection{Previous chatbot solution details}

The Data and Artificial Intelligence Office (DAIO) within the Governance Areas of Intesa Sanpaolo has realized a minimum viable product (MVP) to implement a Generative AI-based virtual assistant to replicate across various domains, with the initial release focused on the HR domain. The MVP aims to address a problem that is both simple and complex. It is simple because the literature and technological solutions for a chatbot are well-established, making it a straightforward initial step. However, the complexity arises when considering advanced functionalities, which significantly increase the scenario's complexity. For example, consider requests that require handling across multiple domains, which necessitates implementing numerous procedural software functionalities within the chatbot. 

An employee of Intesa Sanpaolo can interact with the chatbot and discuss topics based on the HR domain knowledge base. The platform supports defining responses based on accessible user documentation. Additionally, the system offers direct visibility into the sources used to generate responses, along with performance monitoring functionalities to assess the chatbot's effectiveness and key performance indicators. The chatbot is composed of two main components: an \textit{AI Web Portal} and an \textit{AI Backend}. The AI Backend utilizes several cloud functionalities and LLM services. Periodically, the AI application is called upon by the HR domain to ingest new data useful for the chatbot. The ingestion pipelines are capable of managing different types of content, such as text, infographics, and markup languages. Another functionality acquired by the system is the capability to apply access control policies to the knowledge base (KB). In fact, the entire KB is tagged with a proper role that represents a boolean condition over some attributes that the user has. In the end, the system has a suite of tools for monitoring the performance of the chatbot and the efficiency of the responses based on \cite{es2023ragas}.

\label{appendix_a1}
\subsection{Facilitator and Mediator Illustration}

\begin{figure*}[ht]
  \centering
  \includegraphics[width=350pt]{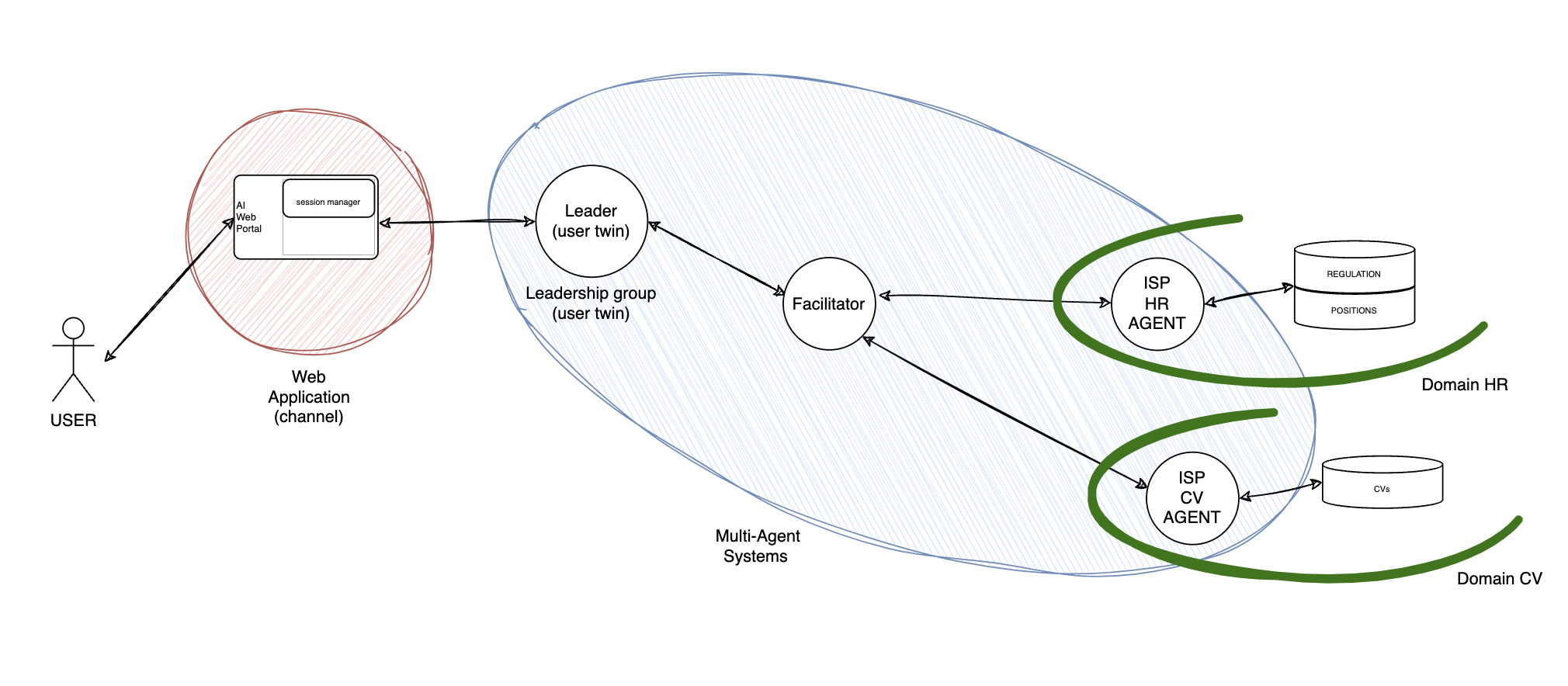}
  \caption{
  The figure illustrates the following scenario: there are two domains—the HR domain and the CV domain. In the HR domain, we find regulations and available positions within the company. In the CV domain, we find candidate CVs for the company. In the multi-agent system, there is a digital twin of the user, a facilitator, and two agents that can operate within these domains. In this simulation, the agents can access the knowledge of their respective domains.
  }
  \label{fig:facilitator}
\end{figure*}

\begin{figure*}[ht]
  \centering
  \includegraphics[width=350pt]{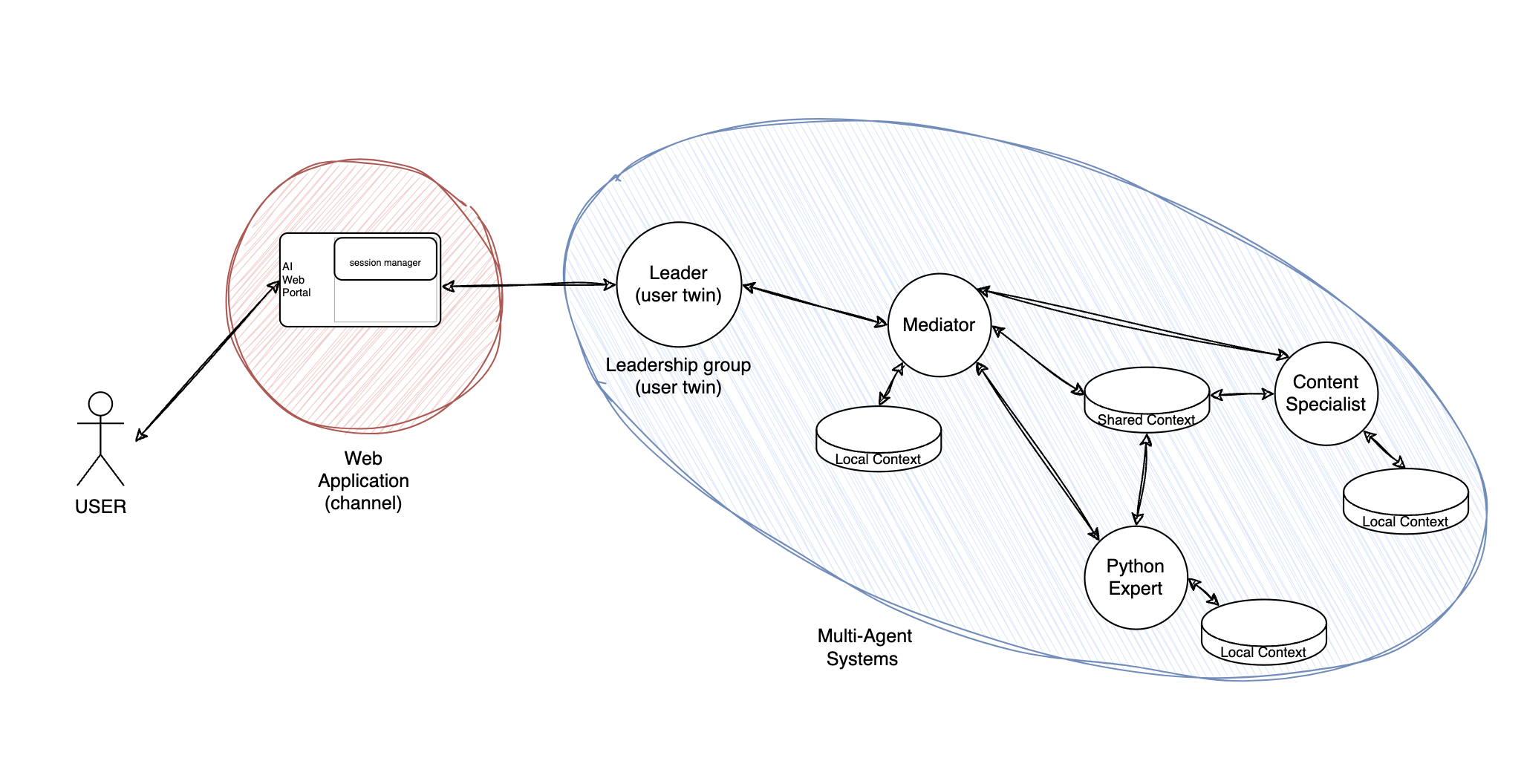}
  \caption{
  The figure illustrates the following scenario: a mediator agent creates two agents to solve a problem. As described in section \ref{fsm-mediator}, the mediator follows different stages to solve the requested problem with the other agents.
  }
  \label{fig:mediator}
\end{figure*}

Figures \ref{fig:facilitator} and \ref{fig:mediator} illustrate a specific case for the facilitator and the mediator, respectively. These simulations showcase the MAS's capability to manage multiple domains by demonstrating how the facilitator agent fuses knowledge from various domains and how the mediator agent engages in decision-making processes. Each figure provides a visual representation of the system's operational effectiveness in handling complex tasks within the system.

\label{appendix_a3}
\subsection{Sample Application}
\label{appendix_b1}

In the repository \href{https://github.com/2mmanu/henry}{github.com/2mmanu/henry}, you’ll find a customizable Kubernetes application designed for deploying a MAS based on HEnRY. In this case, the application is packaged as a Helm chart. Helm is a package manager for Kubernetes that simplifies the deployment and management of applications. The Helm chart in this repository includes the necessary configurations to deploy the HEnRY MAS, which comes with a web app, a digital twin, and a facilitator by default. You can customize the deployment by modifying the Helm chart values as shown in listing \ref{code:values_base}.

To define domain agents within the application, you need to specify a domain group with a suitable name and a list of agents. Each domain agent should have a defined name, some informational details, and a parent. The parent can be either the facilitator or another domain agent acting as a local facilitator within the domain. 

When an agent has one or more children, it takes on the role of facilitator. A facilitator can direct a child agent to address specific problems and provide a syntax for utilizing the child’s capabilities. This syntax is used within the network to handle requests effectively.

\label{code:values_base}
\begin{lstlisting}[
    basicstyle=\ttfamily\footnotesize, % Riduce la dimensione del carattere
    numbers=left,
    numberstyle=\tiny\color{gray},
    stepnumber=1,
    numbersep=10pt,
    gobble=0,
    tabsize=2,
    frame=single,
    showstringspaces=false,
    keywordstyle=\color{blue},
    commentstyle=\color{gray},
    stringstyle=\color{red},
    columns=fullflexible,
    breaklines=true, 
    breakatwhitespace=true,
    caption={Helm chart values},
]
webapp:
  active: true
  vesion: 0.1.1.dev33

twin:
  version: 0.1.1.dev33
  podTemplates:
    replicaCount: 2

facilitators:
  - name: "facilitator"
    podTemplates:
      replicaCount: 1

domains:
  - name: "hr-domain"
    agents:
      - name: "isp-hr-expert"
        parent: "facilitator"
        info:
          agentDescription: |
              HR Assistant provides information regarding salaries, benefits, 
              compensation policies, and other HR-related issues, helping to 
              determine a competitive and appropriate  salary offer. 
          exampleQuestions: | 
              - What is an appropriate starting salary for the candidate?
              - What benefits and extra compensation can the candidate expect?
              - What is the standard salary range for this position in our company?
  - name: "cv-domain"
    agents:
      - name: "isp-cv-expert"
        parent: "facilitator"
        info:
          agentDescription: |
            CV Assistant manages candidates' resumes and provides detailed 
            information about them, such as their work experience, education, 
            and references. 
          exampleQuestions: | 
            - Who is the candidate and do we have their resume? 
            - Can you provide me with a copy of the candidate's resume? 
            - What are the candidate's past work experiences?  
\end{lstlisting}


\end{document}